\documentclass[conference]{IEEEtran}
\IEEEoverridecommandlockouts
\usepackage{cite}
\usepackage{amsmath,amssymb,amsfonts}
\usepackage{algorithmic}
\usepackage{graphicx}
\usepackage{textcomp}
\usepackage{xcolor}
\def\BibTeX{{\rm B\kern-.05em{\sc i\kern-.025em b}\kern-.08em
    T\kern-.1667em\lower.7ex\hbox{E}\kern-.125emX}}
\begin{document}

\title{Beam Test Results of the RADiCAL - a Radiation Hard Innovative EM Calorimeter\\
%{\footnotesize \textsuperscript{*}Note: Sub-titles are not captured in Xplore and
%should not be used}
\thanks{We thank the staff at the Fermilab Test Beam Facility (FTBF). This work has been supported in part by: DOE grant DE-SC0017810, NSF grant NSF-PHY-1914059, the University of Notre Dame – through the Resilience and Recovery Grant Program, and QuarkNet for HS Teacher and Student support.}
}

\author
{
	\IEEEauthorblockN
	{
		James Wetzel\IEEEauthorrefmark{1},
		Dylan Blend\IEEEauthorrefmark{1},
		Paul Debbins\IEEEauthorrefmark{1},
		Max Hermann\IEEEauthorrefmark{1},
		Ohannes Kamer Koseyan\IEEEauthorrefmark{1},
		Gurkan Kamaran\IEEEauthorrefmark{1},\\
		Yasar Onel\IEEEauthorrefmark{1},
		Thomas Anderson\IEEEauthorrefmark{2},
		Nehal Chigurupati\IEEEauthorrefmark{2},
		Brad Cox\IEEEauthorrefmark{2},
		Max Dubnowski\IEEEauthorrefmark{2},
		Alexander Ledovskoy\IEEEauthorrefmark{2},\\
		Carlos Perez-Lara\IEEEauthorrefmark{2},
		Thomas Barbera\IEEEauthorrefmark{3},
		Nilay Bostan\IEEEauthorrefmark{3},
		Kiva Ford\IEEEauthorrefmark{3},
		Colin Jessop\IEEEauthorrefmark{3},
		Randal Ruchti\IEEEauthorrefmark{3},\\
		Daniel Ruggiero\IEEEauthorrefmark{3},
		Daniel Smith\IEEEauthorrefmark{3},
		Mark Vigneault\IEEEauthorrefmark{3},
		Yuyi Wan\IEEEauthorrefmark{3},
		Mitchell Wayne\IEEEauthorrefmark{3},
		Chen Hu\IEEEauthorrefmark{4},\\
		Liyuan Zhang\IEEEauthorrefmark{4},
		Ren-Yuan Zhu\IEEEauthorrefmark{4},
	}\\
	
	\IEEEauthorblockA
	{
		\parbox[t]{1.9in}
		{
			\centering \normalsize
			\IEEEauthorrefmark{1}
			\textit{Dept. of Physics \& Astronomy} \\
			\textit{The University of Iowa}\\
			Iowa City, USA
		}
		\parbox[t]{1.8in}
		{
			\centering \normalsize
			\IEEEauthorrefmark{2}
			\textit{Dept. of Physics} \\
			\textit{University of Virginia}\\
			Charlottesville, USA
		}
		\parbox[t]{1.9in}
		{
			\centering \normalsize
			\IEEEauthorrefmark{3}
			\textit{Dept. of Physics \& Astronomy} \\
			\textit{University of Notre Dame}\\
			Notre Dame, USA 
		}
		\parbox[t]{1.8in}
		{
			\centering \normalsize
			\IEEEauthorrefmark{4}
			\textit{Dept. of Physics} \\
			\textit{Caltech}\\
			Pasadena, USA 		
		}
	}
}

\maketitle

\begin{abstract}
High performance calorimetry conducted at future hadron colliders, such as the FCC-hh, poses a significant challenge for applying current detector technologies due to unprecedented beam luminosities and radiation fields. Solutions include developing scintillators that are capable of separating events at the sub-fifty picosecond level while also maintaining performance after extreme and constant neutron and ionizing radiation exposure. 

The RADiCAL is an approach that incorporates radiation tolerant materials in a sampling ‘shashlik’ style calorimeter configuration, using quartz capillaries filled with organic liquid or polymer-based wavelength shifters embedded in layers of tungsten plates and LYSO crystals.

This novel design intends to address the Priority Research Directions (PRD) for calorimetry listed in the DOE Basic Research Needs (BRN) workshop for HEP Instrumentation. Here we report preliminary results from an experimental run at the Fermilab Test Beam Facility in June 2022.  These tests demonstrate that the RADiCAL concept is capable of $\text{\textless}$ 50 ps timing resolution.
\end{abstract}


\begin{figure}[ht]
	\centering
	\includegraphics[width=0.45\textwidth]{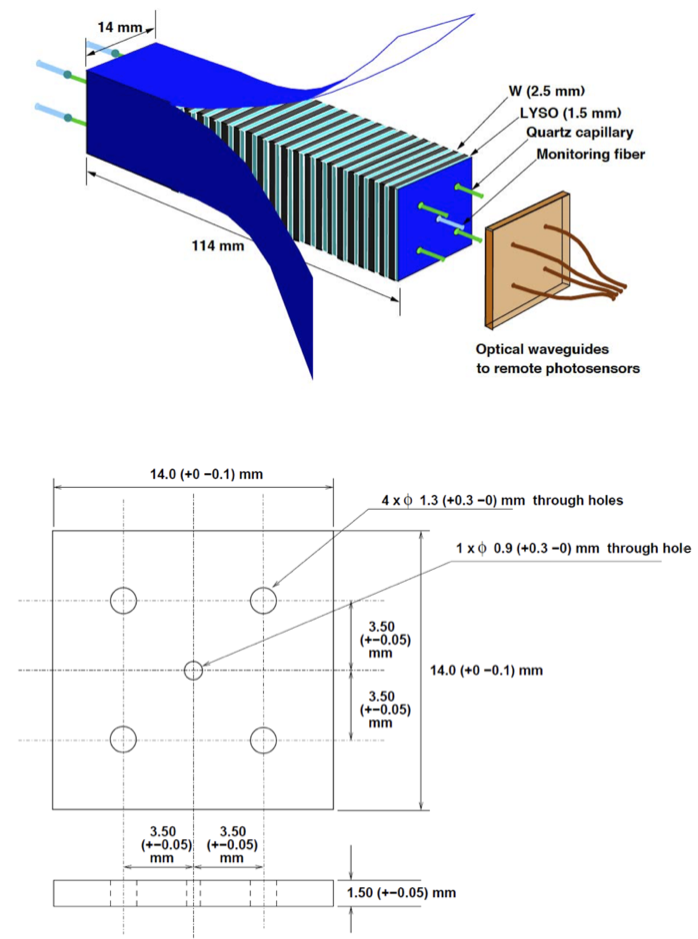}
	\caption{Top: Diagram of a RADiCAL module. Bottom: Diagram of a RADiCAL tile.}
	\label{fig_rad}
\end{figure}

\section{Introduction}

The RADiCAL (radiation-hard innovative calorimeter) was initially conceived to target precision calorimetry for use in future hadron collider experiments \cite{cpad}.  The requirements of such a collider include timing resolution under 50 ps and an EM energy resolution approaching $\sigma_E/E = 10\%\sqrt(E) \oplus 0.3/E \oplus 0.7\%$ \cite{fcc}.

To address these challenges the RADiCAL concept envisions a shashlik style sampling calorimeter composed of many individual modules each 13 cm long and 1.4 x 1.4 cm$^2$ in cross section. Fig. \ref{fig_rad} is a generic schematic of a single prototype RADiCAL module. Comprising such a module are 28 tiles of tungsten (2.5 mm thick) and 29 tiles of LYSO (1.5 mm thick). The radiation length of this structure is 4.7 mm and the Moliére radius is 13.7 mm \cite{rad2}. There are four 1.3 mm and one 0.9 mm diameter holes in each tile to allow for the insertion of quartz capillaries through the length of the module for optical readout, Fig. \ref{fig_rad}, Bottom. 

Monte Carlo simulations using GEANT4 \cite{geant41,geant42,geant43} and previous test beam experiments of the RADiCAL concept using a 4 x 4 array of RADiCAL modules demonstrated realistic expectations for achieving energy resolution performance goals \cite{mdpi,rad2}, and that testing an individual module is sufficient to investigate timing characteristics \cite{rad2}.

To further explore the potential timing characteristics, a single RADiCAL module was tested at the Fermilab Test Beam Facility (FTBF)  \cite{FTBF} in June 2022 using a low energy mixed pion-electron beam E $=$ 28 GeV, with preliminary results reported here.

\section{Experimental Setup}

\begin{figure}
	\centering
	\includegraphics[width=0.35\textwidth]{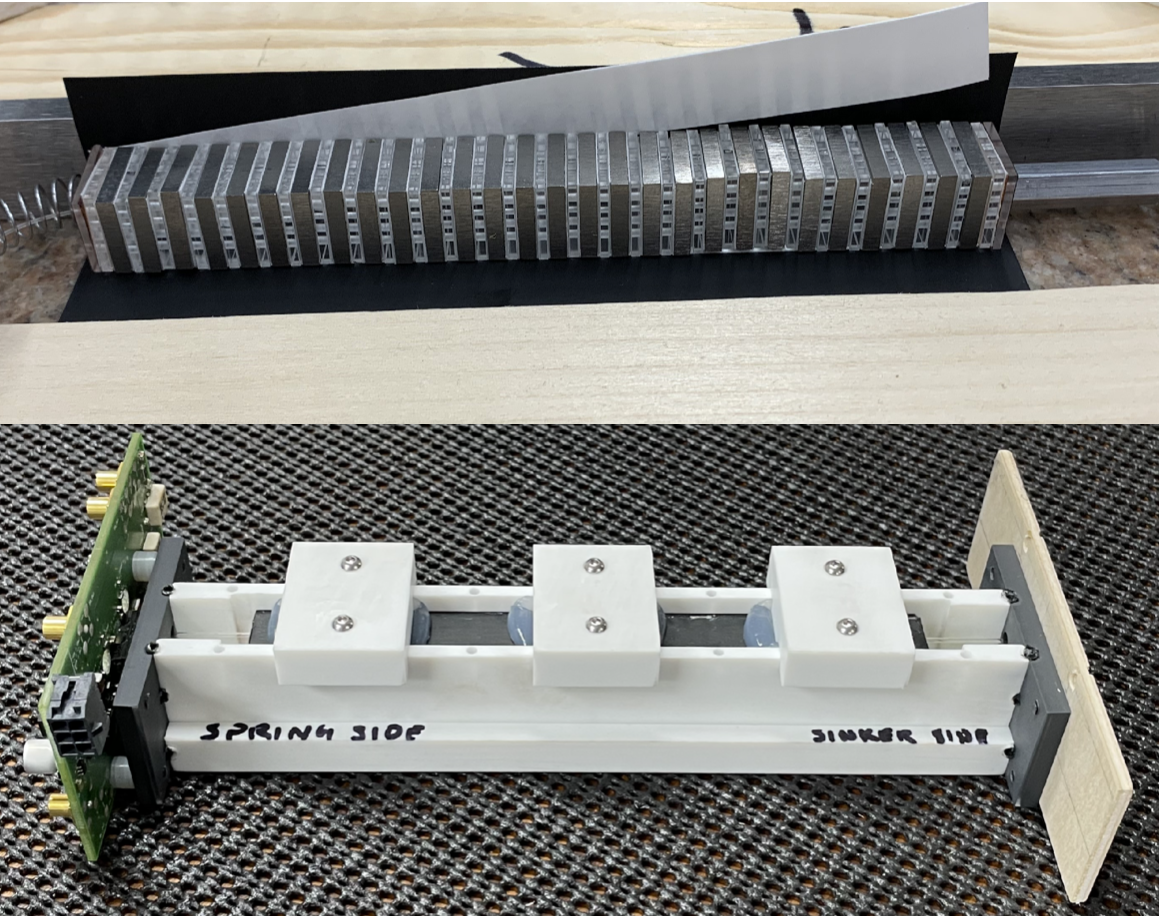}
	\caption{Assembly of a RADiCAL module. The top shows the tiles being stacked, the bottom shows the delrin assembly.}
	\label{fig_ass}
\end{figure}

\begin{figure}
	\centering
	\includegraphics[width=0.4\textwidth]{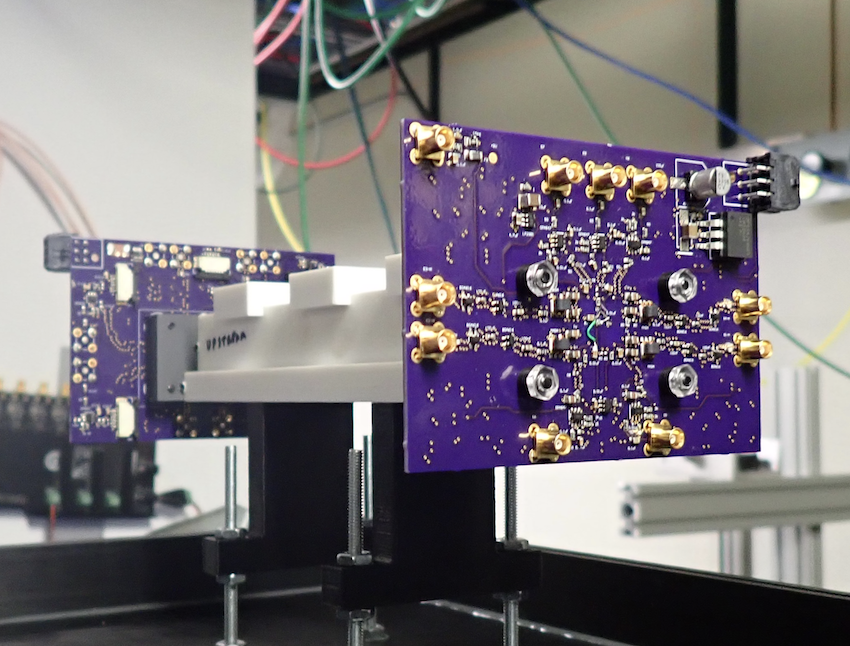}
	\caption{A RADiCAL module in the beam line at Fermilab Test Beam Facility, June, 2022.}
	\label{fig_rad2}
\end{figure}

A single RADiCAL module was prepared by stacking alternating tiles of LYSO and tungsten, separated by laser-cut sheets of Tyvek, and placed in a milled delrin housing, shown in Fig. \ref{fig_ass}. Quartz capillaries were then inserted and readout cards attached. A RADiCAL module with both electronics cards attached is shown in the beam line in Fig. \ref{fig_rad2}.

\begin{figure}[ht]
	\centering
	\includegraphics[width=0.3\textwidth]{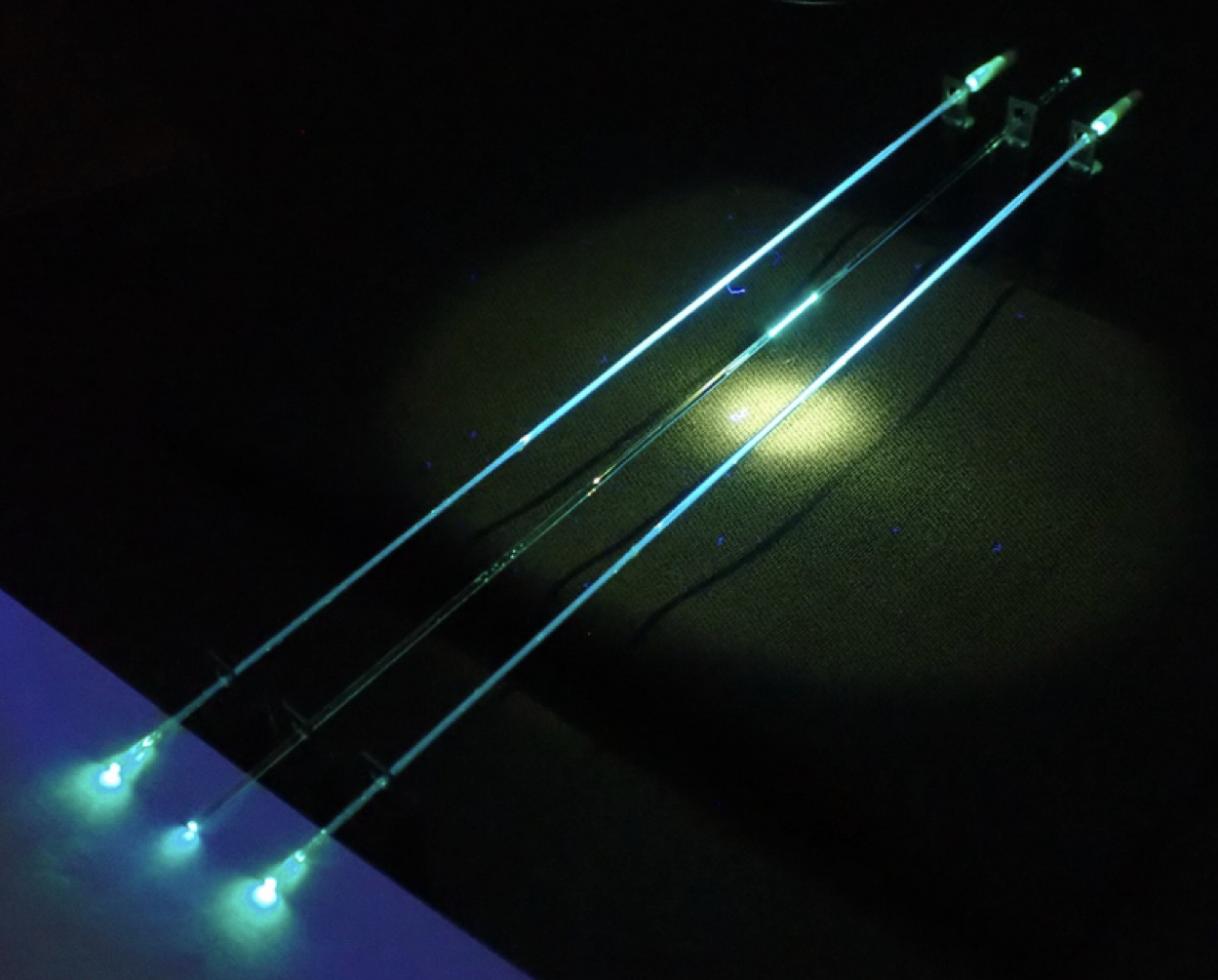}
	\caption{A timing capillary between two energy capillaries.}
	\label{fig_caps}
\end{figure}

\begin{figure}
	\centering
	\includegraphics[width=0.49\textwidth]{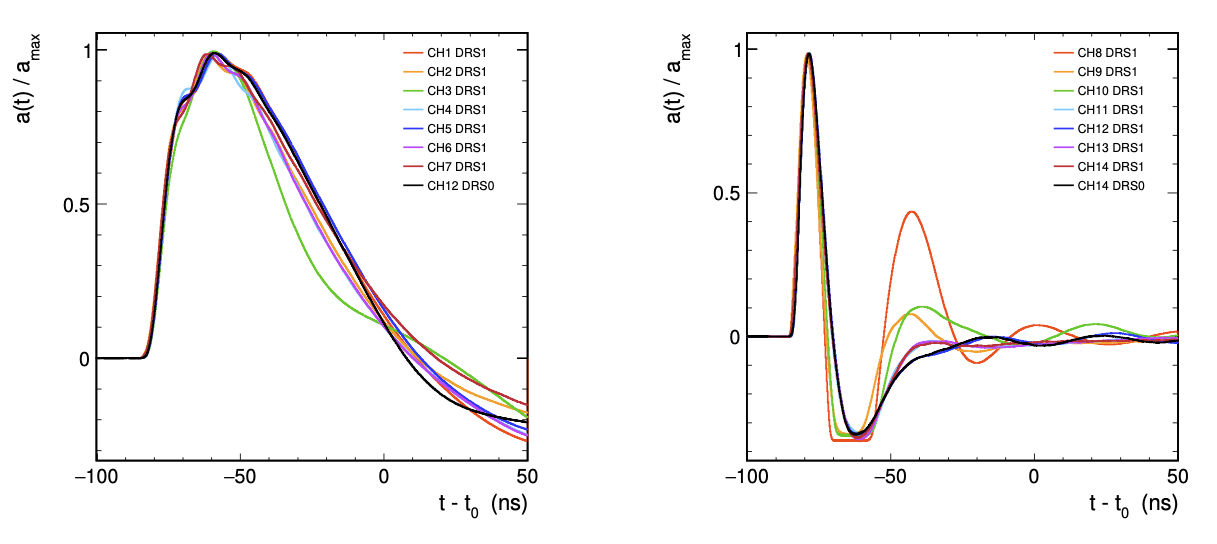}
	\caption{Left: Average normalized waveforms seen by the low gain). Right: high gain channels during the June, 2022 run.}
	\label{fig_waves}
\end{figure}

A RADiCAL capillary is a hollow quartz tube. Each capillary is 18 cm long, covering the active volume of the module and extending 2.5 cm from the module on either end, delivering light from the module to silicon photomultiplers (SiPMs) mounted on readout cards.

Two varieties have been used in tests of the RADiCAL concept. One is referred to as an energy capillary, and one is referred to as a timing capillary.

Energy capillaries are 1000 $\mu$m in diameter, with a 400 $\mu$m bore filled with the organic liquid EJ309 doped with proprietary wavelength shifter DSB1 \cite{eljen}. One end of each liquid filled capillary has its core plugged and fused shut with a ruby quartz filament of 5mm length. The ruby absorbs wave-shifted light traveling principally through the liquid core, preventing it from reaching the photosensors, while allowing light traveling preferentially through the core and quartz walls to be detected. This technique levels and makes uniform the longitudinal response of wave-shifted light collection \cite{rad2}.

Timing capillaries have an outer diameter of 1150$\mu$m and an inner bore with 950 $\mu$m diameter and contain no liquid. In each end, the bore is plugged with a solid quartz rod and fused to create a single element. The two rods leave a gap at the location of shower max within the RADiCAL, determined as mentioned earlier from GEANT4 simulation \cite{rad2}, and that gap is filled with a 15mm long polymer fiber containing DSB1 waveshifter of 900 $\mu$m diameter. The light generated within a timing capillary therefore occurs from a region about shower max. 

Fig. \ref{fig_caps} reveals this localized light generation, comparing the UV illumination of a timing capillary situated between two energy capillaries in this figure.

These capillaries can be in any configuration within the module. The June 2022 configuration reported here used four timing capillaries. The center hole (envisioned for calibrating multiple modules) was unused in this test.

\begin{figure*}[ht]
	\centering
	\includegraphics[width=0.8\textwidth]{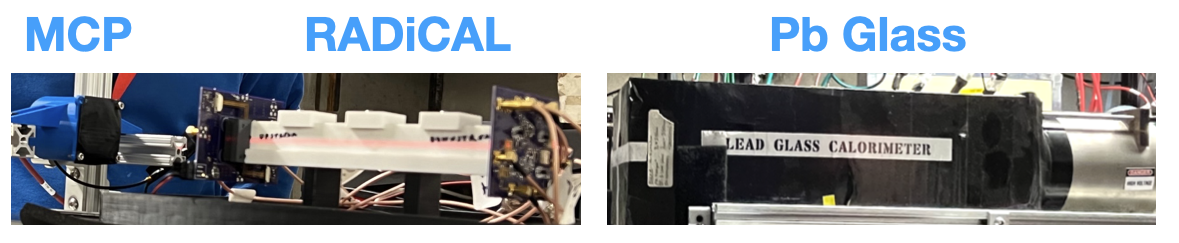}
	\caption{The MCP, the RADiCAL, and the LGC in alignment at the Fermilab Test Beam Facility. The beam travels from the left to the right, hitting the MCP first.}
	\label{fig_setup}
\end{figure*}

\begin{figure*}[ht]
	\centering
	\includegraphics[width=0.8\textwidth]{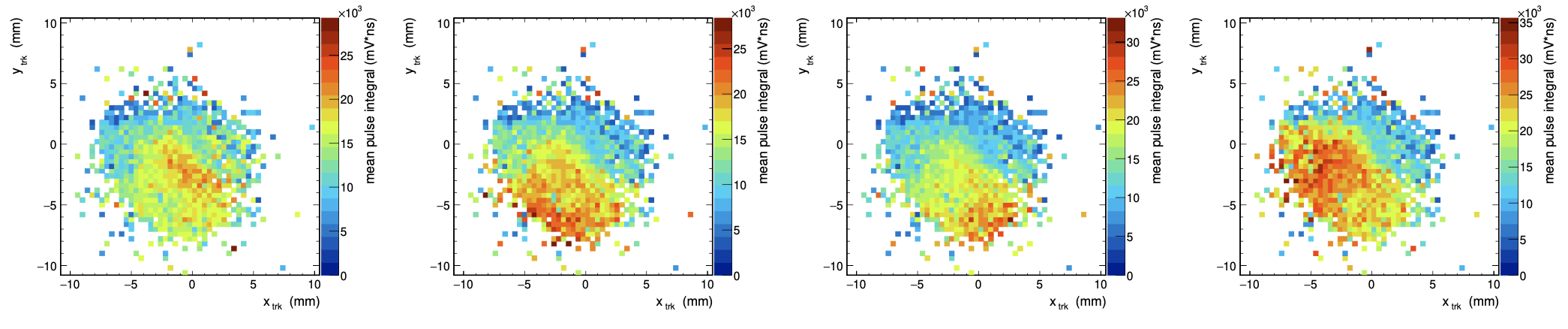}
	\caption{Integrated low-gain signals in each capillary vs location of the particle in X and Y, demonstrating that larger signals are seen when particles hit closer to a capillary.}
	\label{fig_pos2}
\end{figure*}

When particles enter the RADiCAL, they cause the LYSO tiles to scintillate. This light is absorbed by the DSB1 in the timing capillaries, waveshifted, then totally internally reflected to both ends where the capillaries contact the faces of the SiPMs. 

Each SiPM had two readout circuits, one with a low gain amplifier and one with a high gain differential amplifier \cite{stefan}. The low gain amplifier was intended to give clear pulse shapes for energy measurement. The high gain amplifier was intended to give fast sharp pulses for precise signal arrival time measurements. Fig. \ref{fig_waves} shows average waveforms from the June 2022 run.

Fig. \ref{fig_setup} shows the experiment's three important elements. The RADiCAL module was placed in the MT6 beam at the FTBF and aligned with a Hamamatsu R3809U-50 MCP to act as the trigger for data acquisition. This MCP has an expected time resolution $\sigma \approx$ 10 ps \cite{mcp}. Behind these two modules was positioned a lead glass calorimeter (LGC) of dimensions 15 x 15 x 50 cm$^3$. Its purpose was threefold: to record the energy of electromagnetic showers that might be only partially contained within the RADiCAL module; to measure the energy of beam particles that miss the RADiCAL altogether (including those particles which pass through the un-instrumented central hole of the RADiCAL); and to identify and discard hadrons and minimum ionizing particles (MIPs) that pass directly through the RADiCAL. This latter feature was important as the 28 GeV beam at the FTBF consists of electrons with a comparable admixture of pions and minimum ionizing particles or MIPs, \cite{FTBF}.

Additionally, a silicon telescope \cite{pixel} was positioned upstream of these elements in the beam line, to provide tracking information for incoming beam particles and to measure their incident positions on the RADiCAL module.

\begin{figure}
	\centering
	\includegraphics[width=0.4\textwidth]{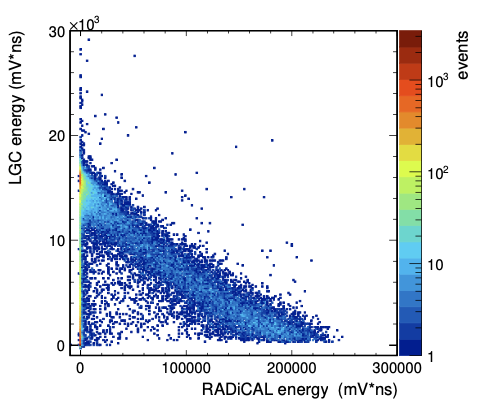}
	\caption{Integrated signals (mV$\cdot$ns) in the LGC vs integrated low gain signals in the RADiCAL module for 28 GeV beam.}
	\label{fig_inten}
\end{figure}

Fig. \ref{fig_pos2} shows the measured energy vs projected track position from the silicon telescope for each capillary, showing a clear position dependence on how much light a capillary detects. This feature can be used to locate a particle (charged or uncharged) within a RADiCAL module based on the relative intensities of each capillary and hence with or without incident beam location information.

Data was acquired using two 16 channel CAEN DT5742 desktop digitizers. The signals from the RADiCAL and the LGC were fed into the digitizers, and the signal from the MCP was fed into the trigger input on both. A particle transiting the MCP creates a signal, triggering the DT5742 to start acquiring data. The MCP has a larger active area than the RADiCAL, so the triggering particle could deposit its energy in either the RADiCAL module or the LGC, or both. This allows the LGC to identify particles that the RADiCAL missed or only partially contained and allows us to select only events that were primarily contained by the RADiCAL.

\begin{figure*}[ht]
	\centering
	\includegraphics[width=0.95\textwidth]{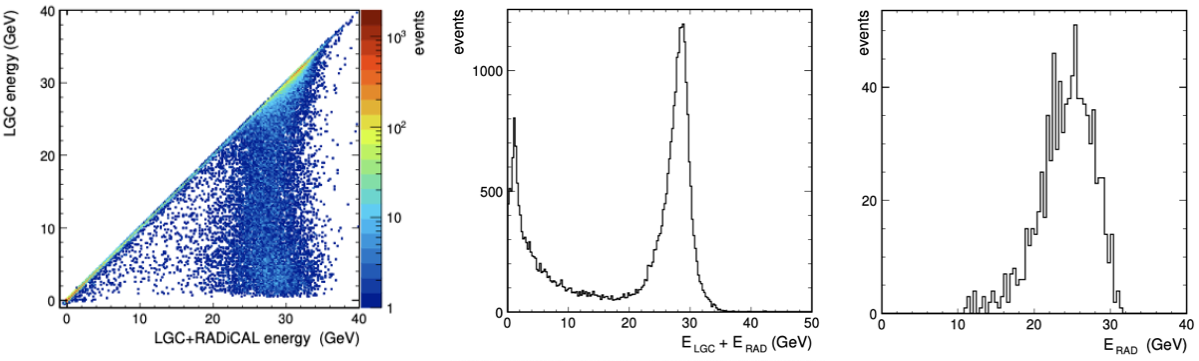}
	\caption{Left: Energy (GeV) in the LGC vs the sum of energy in the LGC and the RADiCAL module. Center: Beam energy as reconstructed from the sum of LGC and RADiCAL signals with electron peak, MIP peak, and hadron tail present. Right: Events reconstructed by the RADiCAL and the LGC with 23 $<$ E$_{LGC}$+E$_{RAD}$ $<$ 33 GeV, event track within 3 mm of the RADiCAL center, and E$_{RAD}$ $>$ 10 GeV.}
	\label{fig_energy2}
\end{figure*}

The data recorded online were converted from a binary format to the ROOT format \cite{root}, the framework within which the data was subsequently analyzed for this report. Approximately 70,000 events were taken with the 28 GeV beam in June, 2022.

\section{Analysis}

All waveforms were pedestal-subtracted with pedestal determined by averaging the first $\approx$ 25 ns of an event, or approximately 125 samples. Samples are taken every 0.2 ns. Low gain signals were fit to determine the signal integral in units of mV$\cdot$ns.

To reconstruct the incident particle energy in GeV, the integrated signal in the LGC was plotted against the sum of all eight integrated low gain signals from the RADiCAL module, in Fig. \ref{fig_inten}. The signals in mV$\cdot$ns for each event in each detector were then scaled to match the nominal 28 GeV beam energy.  This is shown graphically in Fig. \ref{fig_energy2}, Center.

Events were then selected for timing analysis based on three main criteria: To select electron candidates from the mixed beam, the combined signal in the LGC and the RADiCAL was in the range 23 to 33 GeV. To discard tracks through the un-instrumented 1mm diameter central hole of the RADiCAL, the signal in the RADiCAL module was required to be greater than 10 GeV. To minimize the position dependence of the timing measurement, the location of the incident track position determined by the silicon telescope fell within a circle of 6 mm diameter at the center of the RADiCAL module, as there were not enough statistics to correct for position dependence. We opted to select events from a small region in the center where we can assume the position dependence is negligible. The reconstructed energies of these selected events are shown in Fig. \ref{fig_energy2}, Right.

After this event selection, the time stamps were established for all eight high gain readout channels as follows: a timestamp was set the moment a high gain signal crossed a specific threshold. The level of this threshold was optimized for each channel to achieve the best time resolution (see Fig. \ref{fig_waves}, Right). Then, event by event, the average of these eight time stamps was then compared with the corresponding reference time stamp provided by the trigger MCP. The resolution of this timing difference $\sigma$ (t$_{RAD}$ – t$_{MCP}$) was then plotted as a function of the electron energy detected in the RADiCAL module, Fig. \ref{fig_timing}.

\section{Results}

\begin{figure}
	\centering
	\includegraphics[width=0.35\textwidth]{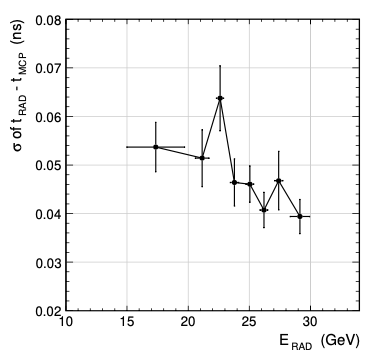}
	\caption{Average timing resolution vs energy of the RADiCAL module.}
	\label{fig_timing}
\end{figure}

As Fig. \ref{fig_timing} indicates, the timing resolution of the RADiCAL module $\sigma$ (t$_{RAD}$ – t$_{MCP}$) improves monotonically with increasing detected energy. This result is consistent with expectations from previous studies, which showed improved timing resolution with increasing signal amplitude \cite{rad2, born}. Averaged over the full detected electron energy range: E $>$ 15 GeV, the resolution of the timing difference is: $\sigma$ (t$_{RAD}$ – t$_{MCP}$) = 49.5ps $\pm$ 5 ps, demonstrating that the RADiCAL can achieve its time resolution objective.

In the portion of the data near the peak of the detected electron energy spectrum: 23 $<$ E$_{RAD}$ $<$ 30 GeV, the resolution of the timing difference is: $\sigma$ (t$_{RAD}$ – t$_{MCP}$)  =  43 ps $\pm$ 4.4 ps.  Given that the timing resolution of the MCP tube is $\sigma$ (t$_{MCP}$) $\approx$ 10 ps, this yields a measured value of:  $\sigma$ (t$_{RAD}$) = 42 ps $\pm$ 4.3 ps for a detected particle energy of E $\approx$ 28 GeV.

This preliminary result will be further refined with improved particle tracking while probing the timing performance as a function of detected electron energy.

\section{Conclusion}

A single module of a shashlik style sampling calorimeter (the RADiCAL) comprised of alternating layers of LYSO and W, embedded with quartz capillaries filled with wavelength shifters were tested at Fermilab in June 2022. This test has demonstrated that the RADiCAL concept can achieve sub 50 ps timing resolution for a detected electron energy E = 28 GeV. Combined with earlier measurements, the RADiCAL has proven itself to be a fast, radiation hard calorimeter concept.

Future plans include measurements of the timing performance over an extended range of electron beam energies from 25 $\textless$ E $\textless$ 200 GeV.

\end{document}